\documentclass[reprint,amsmath,amssymb,aps,floatfix,prl,superscriptaddress]{revtex4-1}

\usepackage{graphicx}
\usepackage{dcolumn}
\usepackage{bm}

\begin{document}

\preprint{}

\title{Crystallization dynamics of a single layer complex plasma}

\author{Peter Hartmann}
\affiliation{Research Institute for Solid State Physics and Optics of the Hungarian Academy of Sciences, P.O.B. 49, H-1525 Budapest, Hungary}

\author{Angela Douglass}
\author{Jorge C. Reyes}
\author{Lorin~S.~Matthews}
\author{Truell W. Hyde}
\affiliation{Center for Astrophysics, Space Physics and Engineering Research (CASPER), One Bear Place 97310, Baylor University, Waco, TX 76798, USA}

\author{Anik\'o Kov\'acs}
\author{Zolt\'an Donk\'o}
\affiliation{Research Institute for Solid State Physics and Optics of the Hungarian Academy of Sciences, P.O.B. 49, H-1525 Budapest, Hungary}

\date{\today}

\begin{abstract}
We report a series of complex (dusty) plasma experiments, aimed at the study of the detailed time evolution of the re-crystallisation process following a rapid quench of a two dimensional dust liquid. The experiments were accompanied by large-scale (million particle) molecular dynamics simulations, assuming Yukawa type inter-particle interaction. Both experiment and simulation show a $\propto t^\alpha$ (power law) dependence of the linear crystallite domain size as measured by the bond-order correlation length, translational correlation length, dislocation (defect) density, and a direct size measurement algorithm. The results show two stages of order formation: on short time-scales individual particle motion dominates; this is a fast process characterized by $\alpha=0.93\pm0.1$. At longer time-scales, small crystallites undergo collective rearrangement, merging into bigger ones, resulting in a smaller exponent $\alpha=0.38\pm0.06$.
\end{abstract}


\maketitle

From the very beginning of laboratory complex (dusty) plasma research, one of the main motivating and promising features of these strongly coupled many-particle systems has been the possibility for modeling classical collective phenomena occurring in atomic matter on a size and time scale that allows direct observation at the particle level \cite{BonitzRev}. Experiments addressing e.g. transport phenomena in two dimensions (heat conductivity \cite{heat}, viscosity \cite{visco}, and self-diffusion \cite{diffusion}), dislocation dynamics \cite{DPdisloc}, various melting scenarios \cite{DPmelt1,DPmelt2,DPmelt3,NosMelt}, and some aspects of freezing \cite{DPfront,Knapek07} have already been carried out. 

The fact that a two-dimensional, hexagonal crystalline structure could easily develop in a laboratory complex plasma experiment was clear from the early days of the field \cite{DPfirst}, even though the theoretical background explaining crystallization in low dimensions \cite{Strand} is still incomplete and the subject of ongoing debate \cite{Gasser}.

Here we study the phenomena of pattern formation occurring when a liquid is rapidly quenched to a solid. Our qualitative expectation is that the initial amorphous liquid structure should evolve toward a ground state crystal through the process of domain coarsening, i.e., merging initially formed small crystallites into fewer and bigger ones. To characterize this process we measured the average linear size of the crystallites and recorded their time evolution. Size measurement is based on: (i) the bond-angular correlation length, (ii) the translational correlation length, (iii) the inverse defect fraction, and (iv) a direct size measurement algorithm similar to the standard ``flood fill'' method used in simple graphical tools.

Ref.~\cite{Harrison} reported a time dependence of the orientational correlation length in the form $t^\alpha$ with $\alpha \approx 1/4$ measured in a two dimensional temperature quenched experiment in a single layer of spherical block copolymer microdomains in a thin film. Recent experiments with superparamagnetic particles ($\propto 1/r^3$ interaction) have shed new light on the dynamics of crystallite formation in colloidal systems resulting in a currently accepted value of $\alpha \approx 0.3$ \cite{Keim_quench}. Recent equilibrium melting studies, on the other hand, expose an important difference between colloidal suspensions and dusty plasmas: dipole systems (with $\propto 1/r^3$ pair-potential) seem to follow the ``dislocation unbinding melting'' (KTHNY) picture \cite{Strand}, while in complex plasmas (with $\propto \exp(-r/\lambda_\text{D})/r$ interaction) rapid domain wall formation is observed, as described by the ``grain-boundary melting'' scenario \cite{NosMelt}.

Various numerical methods (including Monte Carlo, molecular dynamics, dislocation dynamics \cite{Bako07} and phase-field crystal simulations) were applied showing power-law type time evolution of the domain size, as reported in \cite{Huse86} and \cite{Haataja05} with $\alpha \approx 1/3$ for spinodal decomposition, in \cite{Bhat98} with $\alpha \approx 1/3$ for a polymer solution, in \cite{Coveney96} with $\alpha \approx 0.47$ for a binary solution, in \cite{Vega05} with $\alpha \approx 1/4$ for block copolymers, and in \cite{Toyoki93} with $\alpha \approx 0.42$ for a nematic liquid crystal.
 
We have performed a series of complex plasma experiments using the CASPER dusty plasma experimental setup introduced in detail in \cite{setup}. Mono-disperse melamine-formaldehyde micro-spheres with a diameter $d=6.5~\mu$m~$\pm 1$\% were used. An argon gas discharge was driven using a $P=1.5$~W radio-frequency source at 13.56~MHz at a pressure $p=4.7$~Pa and a constant gas flow of about 10 sccm. A single-layer of dust particles was horizontally illuminated by a red laser diode and the scattered light was captured by a ``FASTCAM-1024PCI model 100K'' fast CCD camera from the top at 125 frames per second with a resolution of $512 \times 512$ pixels and a field of view (FOV) having a side length $L=30$~mm. The vertical extension of the layer was monitored using a side-view camera. Our dust clouds consisted of approximately 2000 particles within the FOV. Without external excitation, the system rapidly settled into a ground state configuration exhibiting three or four large crystal domains (forced by the confinement geometry of circular symmetry). The clouds were very stable with no rotation or any other dynamical frustration observed. Melting of the single-layer was induced by temporarily introducing a $\sim 20$~Hz sinusoidal modulation to the DC self-bias of the powered lower electrode, causing a vertical shaking of the layer, which coupled to the horizontal motion of the particles (as studied in detail in \cite{modecopling}), resulting in the complete elimination of the crystalline order within the system. Rapidly switching off the modulation allowed the system to settle into its vertical equilibrium position within less then $2\nu_\text{dust}^{-1}$, where $\nu_\text{dust}=8.8\pm 0.4$~Hz is the observed dominant dust particle oscillation frequency. Both the melting and crystallization cycles were repeated several times in the experiment and images of the particle suspensions were collected continuously.

\begin{figure}[t!]
\includegraphics[width=0.85\columnwidth]{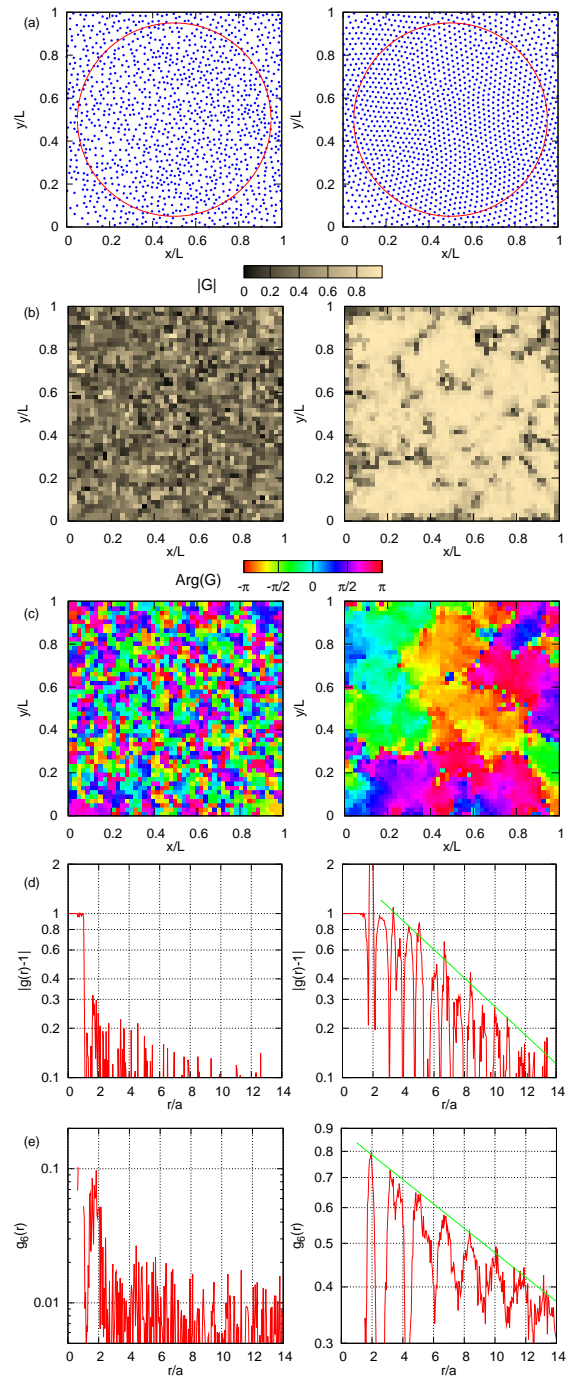}
\caption{\label{fig:pro} 
(color online) Illustration of the data evaluation process for a liquid (left column) and a solid (right column) using snapshot data from experiment. Row (a): particle positions as identified in the raw images. Rows (b) and (c): magnitude and complex phase angle, respectively, of $G_6(r)$ bond order parameter. Rows (d) and (e): $|g(r)-1|$ (pair) and $g_6(r)$ (bond-order) correlation functions of the sample snapshots involving particles situated within the red circles in (a).}
\end{figure}

Figure~\ref{fig:pro} illustrates the principle data evaluation steps for two snapshots, one representative of the system in a liquid phase, the other in a solid configuration. First, the raw images were processed using a particle detection algorithm based on the center-of-mass method introduced and optimized in \cite{PIV}, providing accurate sub-pixel resolution (Fig.~\ref{fig:pro}(a)). Based on these position data, Delaunay triangulation was then performed to find the true nearest neighbors (NN) for every particle in all the snapshots. The defect fraction $D$ (defined as the ratio of particles with a NN number different from 6 and the total particle number) was then determined for every snapshot. Using the neighbor information tables, the complex bond-order parameter
$
G_6(j) = \frac{1}{6} \sum^{\rm N_j}_{l=1} \exp\left[ i 6 \Theta_j(l) 
\right]
$
was calculated for each particle, where $j$ labels the particles, $N_j$ is the number of true nearest neighbors of particle $j$, and $\Theta_j(l)$ is the angle of the $l^{\rm th}$ nearest neighbor bond from an arbitrarily chosen, but fixed direction (the $x$-axis in our case). Based on the spatial distribution of $G_6$, the visualization and identification of the crystallite structure is possible, since the magnitude $|G_6|$ is close to unity inside a grain but small for particles situated within the domain walls (Fig.~\ref{fig:pro}(b)). The complex argument $\text{Arg}(G_6)$, on the other hand, measures the overall angular orientation of a given particle neighborhood, or entire grains (Fig.~\ref{fig:pro}(c)). Both the $g(r)$ pair correlation function and the $g_6(r)=\langle G_6^*(0) G_6(r)\rangle$ bond-order correlation function were calculated for every snapshot. Assuming exponential decay for the envelopes of both $g(r)-1$ and $g_6(r)$ at large $r$, the corresponding correlation lengths $\xi_g$ and $\xi_6$ were then obtained employing an upper envelope least-square fit for each snapshot (Fig.~\ref{fig:pro}(d,e)). To avoid boundary effects, both the correlation lengths and defect fractions were evaluated for particles situated within the red circles shown in Fig.~\ref{fig:pro}(a) (which has a diameter $0.9L$). Distances appear normalized to the 2D Wigner-Seitz radius $a=1/\sqrt{\pi n}$, with $n$ being the surface density of the detected particles.

\begin{figure}[htb]
\includegraphics[width=0.9\columnwidth]{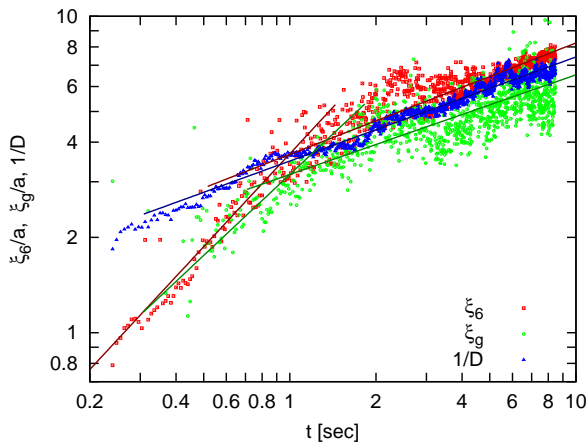}
\caption{\label{fig:corr} 
(color online) Time evolution of the correlation lengths $\xi_6$ (red squares), $\xi_g$ (green circles) and the inverse defect fraction $1/D$ (blue triangles) for a single, representative freezing cycle. Lines represent linear fits on a log-log plot for the short and long time domains.}
\end{figure}

The time-evolution of $\xi_6(t)$, $\xi_g(t)$ and $1/D(t)$ (the inverse defect fraction) is depicted in Fig.~\ref{fig:corr} for a single, representative quenching cycle. One can easily identify two apparently linear regimes on the log-log plot in the case of the correlation functions, showing different time-evolution on the short and long time-scales. Similar observations have been made for colloidal suspensions as reported qualitatively in \cite{Keim_quench}. In our case, the short time period lasts about 10-15 typical oscillation cycles for a single particle located in its equilibrium lattice position with a frequency $\nu_\text{dust}$. The defect density shows, on the other hand, a single characteristic decay time scale. This can be understood realizing, that: (i) at very early times after the temperature drop the particles ballistically move towards their local equilibrium positions, and although crystallites are not yet formed, the defect density is high and its distribution is homogenous. (ii) After some time, small crystallites fill out the space leaving defects to be accumulated in the walls between those grains. In this case (assuming self-similar structure of the domain walls at different times) the relationship between the domain size and the inverse defect fraction can be assumed to be linear. (iii) $G_6$, and thus $g_6(r)$ are continuous quantities, their values can increase for every particle with increasing order. The number of neighbors, on the other hand, is a discrete variable, taking on only integer values. (iv) Finally, the defect fraction ``sensing'' the particles with number of neighbors unequal to six, is less sensitive to the increase of the initially very small bond-angular and positional order at early times. 

Fitting linear functions to the log-log representation of the time-evolution data, as shown in Fig.~\ref{fig:corr}, results in the $\alpha$ exponents. Averaging over 22 individual freezing cycles (of the same particle cloud) results in: $\alpha_6(\text{early})=1.0 \pm 0.16$, $\alpha_6(\text{late})=0.39 \pm 0.1$ for the bond-order correlation length; $\alpha_g(\text{early})=0.87 \pm 0.25$, $\alpha_g(\text{late})=0.34 \pm 0.18$ for the translational correlation length; and $\alpha_D=0.40 \pm 0.05$ for the inverse defect fraction.

Although the $\alpha$ exponents could be derived using the arbitrary units related to the experimental conditions, in order to provide a better understanding of the system, we have derived the following physical quantities based on the dynamical single-particle and collective behavior in the solid phase and reference wave dispersion data from Yukawa lattice calculations \cite{2Dphonon}: (i) Wigner-Seitz radius $a=400\pm15~\mu$m, (ii) Debye screening length $\lambda_\text{D}/a=1.5\pm0.3$, (iii) nominal plasma frequency $\omega_0=55\pm5$~rad/sec, (iv) particle charge $Q/e=10^4\pm 10\%$.

Detailed investigation into the time evolution of the quenching process shows that the dust temperature (measured in terms of kinetic energy) drops from about $200 T_\text{m}$ to $0.4 T_\text{m}$, where $T_\text{m}$ is the critical value associated with the equilibrium solid-liquid transition temperature \cite{22}. This temperature drop can be well fitted with an exponential decay in the form $T(t) \propto  \exp(-t/\tau)$ with $\tau = 0.11\pm 0.01$~s, which is very close to $\nu_\text{dn}^{-1}=0.15\pm0.05$~s the estimated inverse dust-neutral collision frequency. A rapid increase of the observed correlation lengths is initiated immediately when $T$ reaches $T_\text{m} = m\langle v^2\rangle_\text{m}/2k_\text{B}$ in its descent, where $\langle v^2\rangle_\text{m}=\omega_0^2a^2/\Gamma_\text{m}$, with $\Gamma_\text{m}\approx 157$ for Yukawa interaction with $\lambda_\text{D}/a=1.5$ \cite{HPPRE}. The duration of this early period is equal to the time needed for a particle (moving with the observed average velocity) to travel a distance of $\approx 2a$, which is very close to the lattice constant in the final ground state hexagonal configuration, supporting the above described qualitative picture of rapid early ordering induced by single-particle motion.

\begin{figure}[htb]
\includegraphics[width=0.9\columnwidth]{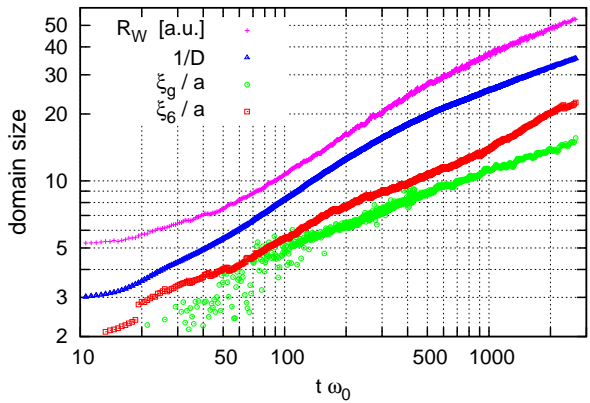}
\caption{\label{fig:sim} 
(color online) Time evolution of the correlation lengths $\xi_6$, $\xi_g$, the inverse defect fraction $1/D$, and the size measured with our ``flood fill'' algorithm from MD simulation. Time is normalized to the 2D nominal plasma frequency $\omega_0=\sqrt{nQ^2/2\varepsilon_0ma}$. Points are shifted for clarity.}
\end{figure}

Our molecular dynamics simulations were carried out for a large 2D particle ensemble (with $N=1~154~000$ particles), interacting through a Yukawa type interaction potential $\Phi \sim \exp(-r/\lambda_\text{D})/r$, where the Debye screening length was chosen to be half the ground state hexagonal lattice constant ($\lambda_\text{D}=0.95 a$).  The topical review article \cite{22} provides an in-depth description of the computational methodology followed in the present work. Quenching in the simulation was achieved by suddenly setting particle velocities to zero and applying a strong frictional force for a short period of time. This allowed changes in the temperature from twice to 0.34 times the melting temperature in less then $10\omega_0^{-1}$ plasma oscillation cycles. Processing of the simulation data was performed in the same manner as previously introduced for the experiment. These results are shown in Fig.~\ref{fig:sim}

To identify the grains, a direct size measuring algorithm similar to the flood fill method was adopted. The technique is very common in graphical applications where it is often used to fill closed areas  with a selected color. In our case we start "coloring" assuming 1600 initial points evenly distributed over our simulation cell. A particle is considered to belong to the same domain as the starting point if (i) it belongs to the neighborhood according to the Delaunay triangulation, and (ii) its complex phase of $G_6$ is within the $\pm 15$ degrees range with respect to that of the starting point. This algorithm results in the number of particles (the weight) of the domain containing the starting point. In the calculation of the average domain weight $\langle W \rangle$ we neglect cases with $W<6$, as these represent cases where the starting point was accidentally picked in a domain wall. The average linear size of the domains is approximated as $R_W \propto \sqrt{\langle W \rangle}$.

Performing linear fitting to the data as shown in Fig.~\ref{fig:sim} for times $t\omega_0 >500$ results in the following exponents: $\alpha_6 \approx 0.41$, $\alpha_g \approx 0.35$, $\alpha_D \approx 0.36$, and $\alpha_W \approx 0.38$. The values obtained by these four independent measures provide consistent results if we assume for the real time evolution of the linear domain size a power law time dependence with an exponent $\alpha = 0.375 \pm 0.03$, corresponding to (and extending) the long time characteristics of the experiment.

In conclusion, we have performed a series of complex plasma experiments measuring the time evolution of the crystallization process of a single layer system after a rapid temperature quench from the liquid to solid state. Our experiments show a rapid early process, driven by single particle motion, where the system undergoes a continuous transition from its initial amorphous state to a polycrystalline state of small domains. After about 10 - 15 particle oscillation cycles, pattern formation slows and further grain coarsening is only possible by collective rearrangement, resulting in the merging of small domains to form larger ones. The linear domain size shows a power-law type time evolution with an exponent $\alpha^\text{early} =0.93 \pm 0.1$ at the short times, while $\alpha^\text{late} =0.38 \pm 0.06$ at long times. The long-time characteristics could be observed under stable temperature conditions, which made the real time-dependence study possible, as it is not altered by systematic changes in the temperature. It reveals the true nature of grain coarsening and pattern formation in 2D, not affected by the often significant and irreproducible influence of the actual conditions in a particular complex plasma experiment \cite{Knapek07}. This result is supported by our molecular dynamics simulation for a 2D Yukawa system resulting in $\alpha^\text{MD} = 0.375 \pm 0.03$. Comparison of our findings reveals satisfactory agreement with earlier results concerning systems with low friction ($0.3 < \alpha < 0.45$) and  with liquid (overdamped) environment ($\alpha < 0.3$) exhibiting a somewhat faster coarsening \cite{Harrison,Knapek07,Harrison,Keim_quench,Huse86,Haataja05,Bhat98,Coveney96,Vega05,Toyoki93}.

\begin{acknowledgments}
The research presented in this letter was supported by the CASPER project, and Hungarian Grants OTKA PD-75113, K-77653, and the J\'anos Bolyai Research Scholarship of the Hungarian Academy of Sciences. The authors are grateful for motivating discussions with G. Gy\"orgyi, I. Groma, P. D. Isp\'anovity, and L. Gr\'an\'asy.
\end{acknowledgments}

%

\end{document}